\documentclass[twocolumn,fleqn,,usenatbib]{mnras}
\usepackage{hyperref}
\usepackage{bm}
\usepackage{multirow}
\usepackage{gensymb}
\usepackage{enumitem}
\usepackage{tabularx}
\usepackage[flushleft]{threeparttable}
\usepackage[]{color}

\usepackage{float}
\usepackage{breqn}
\usepackage{amsmath}

\usepackage{amsfonts,amsmath,amssymb,esint}
\usepackage{graphicx}
\usepackage{color}
\usepackage{xcolor}
\usepackage{url}

\title[Blackbodies Matter]{Blackbodies Matter}

\author[S. Dyda]{
Sergei Dyda,$^{1,2}$\thanks{sdyda@ua.edu}
\\
$^{1}$ Department of Physics \& Astronomy, University of Alabama, Gallalee Hall, 514 University Blvd, Tuscaloosa, AL 35401, USA \\
$^{2}$Department of Astronomy University of Virginia 530 McCormick Rd. Charlottesville, VA 22904, USA
}

\begin{document}

\label{firstpage}
\pagerange{\pageref{firstpage}--\pageref{lastpage}}

\maketitle

\begin{abstract}

We study line driven stellar winds using multifrequency, time-dependent radiation hydrodynamics. We compute the radiation force due to lines, the so called force multiplier, using precomputed photoionization tables and a time-dependent, local SED using a three band approximation within the hydro. We find that accounting for changes in the local SED changes the global properties of the flow. This is due to a combination of a change in the available momentum in the radiation field but also due to changes in the gas/radiation coupling mediated by the force multiplier. The acceleration can be \emph{suppressed}, due to attenuation of UV flux but the force multiplier itself can be \emph{enhanced} due to the overall SED hardening. We do not see evidence of a changing SED leading to formation or growth of azimuthal instabilities, as this effect appears subdominant to other instabilities in the wind. The computational methods presented can be extended to other outflow problems, notably multi-temperature disc winds and shocking in Wolf-Rayet stars.     
\end{abstract}

\section{Introduction}
\label{sec:introduction}

In astrophysical winds, radiation can play a crucial role in driving the outflow of gas. The role of radiation in outflows is two fold. Firstly, the radiation field acts as a source of energy and momentum that may be imparted on the gas to allow it to escape the central objects gravitational well. Secondly, and most crucially, it interacts with the gas microphysically to enhance or suppress the radiation/gas coupling via scattering and absorption opacities and subgrid processes like line emission, recombination and bremsstrahlung. It is ultimately the strength of this radiation/gas coupling that determines what fraction of the radiation fields energy and momentum is available to launch the wind.

Simultaneously accounting for all these physical processes is challenging because of the range of scales involved. Typically, one models the large scale fluid flows using a hydrodynamics code and describes the gas microphysics using a photoionization code. Computational limitations require that the gas microphysics be either approximated by an analytic expression \citep{Blondin1990,Blondin1994}, precomputed over the relevant parameter space \citep{Dyda17} or computed within the hydro but with considerably lower cadence than the hydrodynamics solver \citep{Higginbottom2020}. Alternatively, new approaches have been proposed, using machine learning  on photoionization code data to train a model to predict heating rates \citep{2024arXiv240619446R}. 

Studies have shown in the context of thermally driven winds that the form of the SED can have important effects on the resulting outflow. \cite{Dyda17} performed spherically symmetric simulations of thermally driven outflows where the heating rates were computed from photoionization modeling where the gas was assumed to be irradiated by a variety of composite SEDs. They found that the gross outflow properties (mass flux, outflow velocity, momentum flux) were highly dependent on SED, and not simply bolometric luminosity as might be expected from basic thermal driven wind theory \cite{1999isw..book.....L}. Their method relied on precomputing the heating rates, assuming the entire flow was optically thin and irradiated by the same SED. Similar methods have been used for line driven disc winds, where the flow is assumed to be optically thin and irradiated by a \emph{global} SED \citep{Dannen2023, Dannen2024}. 

Post processing of hydrodynamic outflow models by Monte Carlo radiation transfer codes has shown that this optically thin wind assumption may not be an accurate model for radiation transfer. \cite{Higginbottom14} post processed AGN line driven disc wind simulations \citep{PSK2000} and showed that the ionization structure of the flow was inconsistent with the ionization structure in the hydrodynamics model. The new structure would in fact be \emph{over}-ionized, and the resultant radiation force would be insufficient to launch the outflow. Whether the flow would be able to readjust itself to sustain the launching, such as a failed wind shielding more distant gas, is an open question. The study of such problems could benefit from rapid, accurate, methods which account for a changing SED within the hydrodynamics simulation.  

Typically, the enhancement of the radiation/gas coupling by the ions above that of pure electron scattering is measured in terms of the \emph{force multiplier}
\begin{equation}
    M = \frac{F_{\rm{rad,lines}}}{F_{\rm{rad,es}}},
\end{equation}
which measures the ratio in radiation force due to ionic lines $F_{\rm{rad,lines}}$ to the radiation force due to electron scattering $F_{\rm{rad,es}}$ . Force multipliers greater than order unity allow for winds to be launched from objects where the Eddington parameter
\begin{equation}
    \Gamma \sim \frac{1}{M},
\end{equation}
thus allowing sub-Eddington winds to be radiatively driven if the force multiplier is greater than unity. Typical examples of such objects include O stars where $\Gamma \approx 0.1$, cataclysmic variables where $\Gamma \lesssim 0.01$ and AGN where $\Gamma \lesssim 0.5$.

In a line driven wind, the radiation force is a sum of the contributions due to electron scattering and due to lines, given by
\begin{equation}
F_{\rm{rad}} \propto F_r (1. + M)
\end{equation}
where $F_r$ is the irradiating flux. Changes in the radiative force on the wind are thus due to a combination of 1) a change in flux, whereby less momentum is available to transfer to the wind 2) a change in force multiplier, which changes the radiation/gas coupling. Decreases in flux can occur due to geometric dilution, optical depth effects, scattering or re-emission. Accounting for all but the last effect necessitates the use of some type of radiation  transfer of which several numerical methods have been developed over the years including flux-limited diffusion \citep{1981ApJ...248..321L}, M1 \citep{2007A&A...464..429G}, solving the time-dependent radiation transfer method \citep{Jiang12} or Monte Carlo methods \citep{Matthews2025}. Accounting for the second effect is more challenging, as recomputing the force multiplier within the hydrodynamic simulation is very costly. One possible workaround, as implemented by \cite{Higginbottom2020}, is to rely on an approximation where the radiation transfer is not updated with every hydro step. Most simulations have thus far relied on the assumption that the underlying irradiating SED does not change.  

In this paper, we describe a method to leverage the multifrequency radiation transport of the hydrodynamics code \textsc{Athena++} (\cite{Stone2009}, \cite{Jiang2022}) and the photoionization code \textsc{XSTAR} \citep{XSTAR2001} for optically thick flows in a self consistent way. We perform a series of photoionization simulations, where the gas is irradiated by blackbody spectra of a range of temperatures. From these models, we extract the opacities of the spectral lines due to IR, UV and X-ray bands. This data is stored as a function of the fraction of radiation intensity in the IR, UV and X-ray. Within the hydro simulation, a 3 band approximation is used for the  SED and the resulting force multiplier determined in real time. 

We demonstrate our method by studying 2D, spherically symmetric line driven winds from massive stars. We show that for some winds, changes in the radiation force on the wind are well accounted for from a decrease in radiative flux. However, in other cases, we demonstrate that the radiative force changes due to a change in the force multiplier. The temperature of the blackbody matters in calculating the force multiplier and changes in the radiation force cannot solely be attributed to a loss in available momentum in the radiation field. 

The structure of the paper is as follows. In Section \ref{sec:theory} we describe our method for computing the force multiplier using \textsc{XSTAR} and its implementation in \textsc{Athena++}. We describe our setup for studying winds from massive stars. In Section \ref{sec:results} we describe the results of SED dependent force multipliers on line driven winds. We show explicitly that in some cases, changes in  the wind can be attributed due to a loss of flux, but in others they are due to changes in the force multiplier, which are a consequence of the irradiating SED. We show that accounting for this effect does not introduce additional clump formation, as these effects are subdominant to other instabilities in line driven winds. We conclude in Section \ref{sec:discussion} where we discuss further applications of these methods to other problems of line driven winds. We provide greater details of our method and some benchmarking tests in the appendices.

\section{Simulation Setup}
\label{sec:theory}

\subsection{Photoionization Modeling}
\label{sec:photoionization}

\begin{figure}
    \centering
    \includegraphics[scale=0.55]{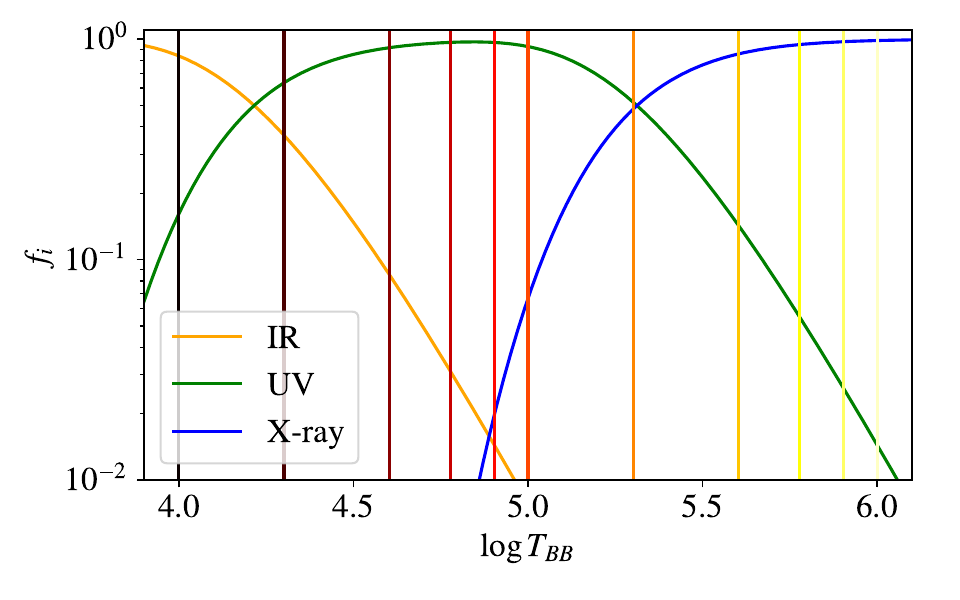}
    \caption{Radiation fraction in IR (orange line), UV (green line) and X-ray (blue line) bands as a function of blackbody temperature. The vertical colored lines indicate temperatures for which we have photoionization modeling. The force multiplier data for these SEDs is shown in Fig \ref{fig:Mt_avg}-\ref{fig:Mt_avg_summary}}
    \label{fig:spectrum}
\end{figure}

We employ the photoionization code \textsc{XSTAR} to compute a grid of models to implement microphysics into our hydro simulations. For a wide range of optical depth parameter $-8 \leq \log t \leq 1$ and photoionization parameter $-2 \leq \log \xi \leq 5$, we compute the force multiplier (see \cite{Dannen19}, hereafter D19 for a full description of our methods) for a gas irradiated by a blackbody spectra at temperature $10^{4} K \leq T_{BB} \leq 10^{6} K$. Notably, D19 used an updated version of \textsc{XSTAR}, including a line list featuring over 2 million lines. We assume solar abundances for atomic species. The radiation force due to spectral lines is characterized by the force multiplier $M$, which is the strength of the radiation force relative to electron scattering. We compiled the contributions to the force multiplier for the IR, UV and X-ray bands ($M_{IR}$, $M_{UV}$ and $M_{X}$ respectively). We defined these for wavelengths $\lambda \geq 7000 \AA$, $200 \AA \leq \lambda \leq 7000 \AA$ and $\lambda \leq 200 \AA$ respectively. A summary of these results is shown in Appendix \ref{sec:FluxAvgMt}.  

In order to implement this grid of models in our hydrodynamics scheme, we compiled the data into tables that could be interpolated over as we ran the hydro simulations. We have previously used this method for incorporating heating and cooling (see for example \cite{Dyda17}) and force multipliers (see for example \cite{Dannen20}. In these previous works, the heating and force multiplier were functions of the ionization state of the gas, temperature or optical depth parameter. Here, the novel aspect is we also include the spectral information of the irradiating SED by indexing the fraction of the radiation energy density in the UV band 
\begin{equation}
    f_{UV} = \frac{J_{UV}}{J}.
\label{eq:fUV}
\end{equation}
Fig \ref{fig:spectrum} shows the radiation fraction in the IR (orange), UV (green) and X-ray (blue) bands as a function of blackbody temperature $T_{BB}$. By breaking up the table into ``cold'' blackbodies with $T_{BB} \leq 6 \times 10^5 $K where $f_{IR} > f_{X}$ and ``hot'' blackbodies with $T_{BB} > 6 \times 10^5 $K where $f_{IR} < f_{X}$ we have two sets of tables which can be characterized by a monotonic parameter $f_{UV}$. This parameter can be computed in the hydro simulation using equation (\ref{eq:fUV}).  

In Appendix \ref{sec:SpecAvg} we describe in more detail our method for interpolating across different values $f_{UV}$ and some tests to check the accuracy of this method. By compiling the force multipliers in this way we can better account for changes in the SED and resultant changes to the force multiplier within the hydrodynamics simulation.

\subsection{Hydrodynamics \& Radiation Transfer}
\label{sec:hydro}

The equations for single fluid,  multi-frequency radiation hydrodynamics are
\begin{subequations}
\begin{equation}
\frac{\partial \rho}{\partial t} + \nabla \cdot \left( \rho \mathbf{v} \right) = 0,
\end{equation}
\begin{equation}
\frac{\partial (\rho \mathbf{v})}{\partial t} + \nabla \cdot \left(\rho \mathbf{vv} + \sf{P} \right) =  \mathbf{G} + \rho \mathbf{g}_{\rm{grav}},
\label{eq:momentum}
\end{equation}
\begin{equation}
\frac{\partial E}{\partial t} + \nabla \cdot \left( (E + P)\mathbf{v} \right) = cG^{0} + \rho \mathbf{v} \cdot \mathbf{g}_{\rm{grav}},
\label{eq:energy}
\end{equation}
\label{eq:hydro}%
\end{subequations}
where $\rho$ is the fluid density, $\mathbf{v}$ the velocity, $\sf{P}$ a diagonal tensor with components $P$ the gas pressure. The total gas energy is $E = \frac{1}{2} \rho |\mathbf{v}|^2 + \mathcal{E}$ where $\mathcal{E} =  P/(\gamma -1)$ is the internal energy and $\gamma$ the gas constant. The gravitational source is due to a star with
\begin{equation}
    \mathbf{g}_{\rm{grav}} = -\frac{GM}{r^2} \hat{r}, 
\end{equation}
where $M$ is the stellar mass and $G$ the gravitational constant. The temperature is $T = (\gamma -1)\mathcal{E}\mu m_{\rm{p}}/\rho k_{\rm{b}}$ where $\mu = 1$ is the mean molecular weight and other symbols have their standard meaning.  

The radiation source terms $\mathbf{G}$ and $cG^0$ are assumed to receive contributions from the continuum and spectral lines
\begin{subequations}
\begin{equation}
\mathbf{G} = \mathbf{G}_{\rm{cont.}} + \mathbf{G}_{\rm{lines}}, 
\end{equation}
\begin{equation}
G^{0} = G^{0}_{\rm{cont.}} + G^{0}_{\rm{lines}}. 
\end{equation}    
\end{subequations}

The continuum radiation field consists of three bands, IR, UV and X-ray and are treated by directly solving the time dependent radiation transport equation using the implicit implementation in \textsc{Athena++} \citep{Jiang2022}.  The radiation transfer equation solved is equivalent to
\begin{dmath}
\frac{\partial I_{\nu}}{\partial t} + c \mathbf{n} \cdot \nabla I_{\nu} = c S_{I,\nu}, 
\label{eq:dIdt}
\end{dmath}
with the source term
\begin{equation}
\begin{aligned}
    S_{I,\nu} = \Gamma^{-3} \rho \Big[ \Big( \kappa_{P,\nu} \frac{c a T^4}{4\pi} &-  \kappa_{E,\nu} J_{\nu,0} \Big) \\
     &- \left( \kappa_{s} + \kappa_{F,\nu} \right) \left(I_{\nu,0} - J_{\nu,0} \right) \Big],
     \end{aligned}
    \label{eq:source}
\end{equation}
where $\kappa_{s}$ is the scattering opacity, $\kappa_{F,\nu}$ is the absorption contribution to the flux mean opacity, $\kappa_{P,\nu}$ the Planck mean and $\kappa_{E,\nu}$ the energy mean opacity. $I_{0,\nu}$ is the intensity in the comoving frame and 
\begin{equation}
    J_{0,\nu} = \frac{1}{4\pi}\int I_{0,\nu} \; d\Omega_0,
\end{equation}
is the corresponding angle averaged comoving frame mean intensity.

With the above assumptions the continuum momentum and energy source terms are then
\begin{equation}
    \mathbf{G}_{\rm{cont.}} = \frac{1}{c} \sum_{\nu} \int \mathbf{n} S_{I,\nu} d \Omega,
\end{equation}
\begin{equation}
    cG^0_{\rm{cont.}} = c \sum_{\nu} \int S_{I,\nu} d \Omega.
\end{equation}

We model the force due to lines via a CAK type prescription using the local continuum flux. Working in the Sobolev approximation, the line force
\begin{equation}
\mathbf{G_{\rm{lines}}} = \frac{\rho \kappa_{es}}{c} \sum_{\nu} \oiint M_{\nu}(t) \mathbf{n} I_{\nu}(\mathbf{n}) d\Omega,
\end{equation}
where the integral is over all radiation rays of the continuum in band $\nu$. The frequency dependent force multiplier is a function of the optical depth parameter 
\begin{equation}
    t = \frac{\rho v_{\rm{th}} \sigma_e}{|dv/dl|},
    \label{eq:dvdl}
\end{equation} 
where $v_{\rm{th}} = 4.2 \times 10^{5} cm/s$ is the gas thermal velocity and $dv/dl$ the velocity gradient along the line of sight of the radiation flux.

The work done by the line force is then
\begin{equation}
G^0_{\rm{lines}}(E) = v \cdot \mathbf{G_{\rm{lines}}}. 
\end{equation}

We express the stellar luminosity via the Eddington parameter 
\begin{equation}
  \Gamma = L_{*} \sigma_e / 4\pi cGM
\end{equation}  
where $L_*$ the stellar luminosity.

\subsection{Simulation Parameters}
We choose parameters relevant to O stars. We take the mass to be $M = 50 M_{\odot}$ and radius $r_* = 10 r_{\odot}$. We set the Eddington fraction $\Gamma = 0.1$. We consider gas temperatures $10^{4} K \leq T \leq 10^{6} K$. For these temperatures, thermal driving is negligible \citep{Stone2009}.  

The simulation region extends radially from $r_* < r < 10 r_*$ and we take a wedge $\Delta \theta = 0.2$ with $N_{\theta} = 64$. This dynamical range allows the wind to fully accelerate and reach its terminal velocity. We use a logarithmically spaced grid of $N_r$ = 1024 points and a scale factor $a_r = 1.008$ that defines the grid spacing recursively via $dr_{n+1} = a_r dr_{n}$. This resolution ensures that we resolve up to the Sobolev length, where our model for the line transfer breaks down.  

At the inner boundary, we impose outflow boundary conditions on $v$ and $E$ while keeping the density fixed at $\rho_* = 10^{-10}$ $g/cm^3$ in the first active zone. This density ensures that we are sufficiently resolving the atmosphere at the base of the wind, but also providing sufficient mass to launch a wind given our choice of Eddington parameter. The continuum intensity of the radially outgoing rays is set by the Eddington parameter while the non-radial rays are set to zero. The input radiation spectra is set by choosing a blackbody temperature $T_{BB}$, which we use to determine the radiation fraction $f_i$ in each band.

At the outer boundary, we impose outflow boundary conditions on $\rho$, $v$ and $E$ and vacuum conditions on the continuum. We use periodic boundaries along the azimuthal boundaries.

\section{Results}
\label{sec:results}

\subsection{Radial Streaming Approximation}

\begin{figure}
    \centering
    \includegraphics[scale=0.52]{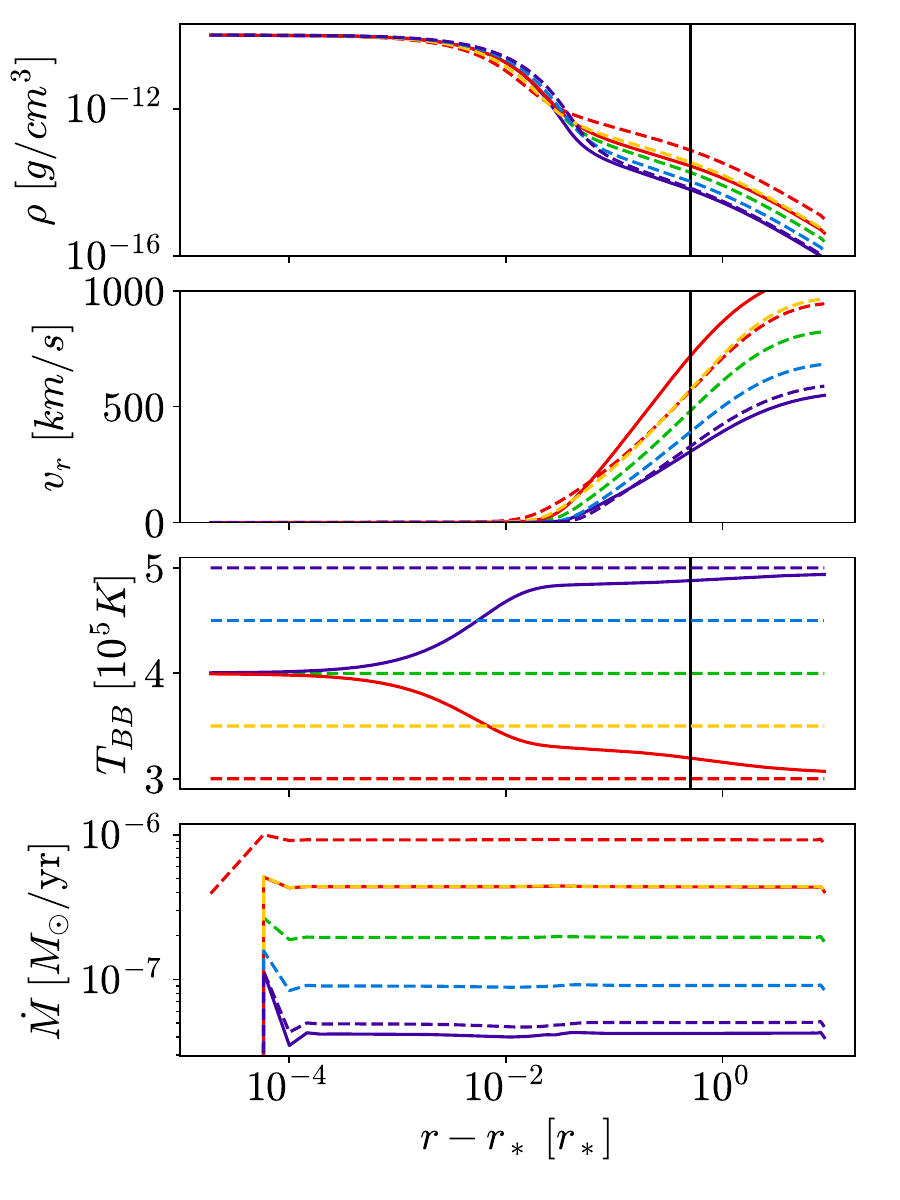}
    \caption{Late time, stationary wind solutions for gas with $\kappa_a = 3 \kappa_{es}$ (solid lines) and $\kappa_a = 0$ (dashed lines). For the optically thin cases, the colors indicate the blackbody temperature of the irradiating spectra with $T_{BB} [K] = 3 \times 10^{5}$ (red), $3.5 \times 10^{5}$ (orange), $4 \times 10^{5}$ (green), $4.5 \times 10^{5}$ (cyan), $5 \times 10^{5}$ (blue). For cases with absorption, the color corresponds to the model with the closest spectra at the outer simulation boundary. The black vertical line indicates the location of the critical point for the $T_{BB} =4 \times 10^{5}$ K model.}
    \label{fig:dynamical_summary}
\end{figure}

\begin{figure}
    \centering
    \includegraphics[scale=0.42]{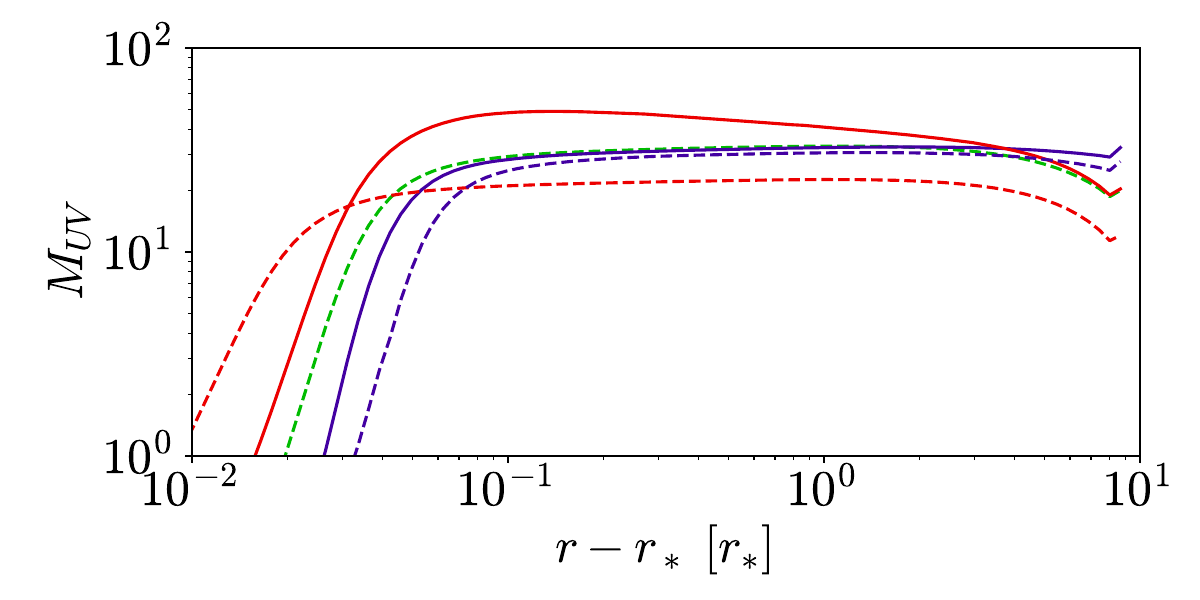}
    \caption{$M_{UV}$ summary for a selection of models shown in Fig \ref{fig:dynamical_summary}. The $T_{BB} =5 \times 10^{5}$ (blue) model have comparable force multipliers, leading to comparable dynamics. In contrast, the $T_{BB} =3 \times 10^{5}$ (red) model has a stronger force multiplier, due to a softening of the irradiating SED. }
    \label{fig:MUV_summary}
\end{figure}

We perform a series of 2D simulations, using the radial streaming approximation where we set $dv/dl \approx dv/dr$ in equation (\ref{eq:dvdl}). This approximation tends to lead to stationary solutions and better approximates a spherically symmetric flow. We therefore restrict our analysis to the radial structure and discuss azimuthal effects in the following section.

As a fiducial model we consider an irradiating blackbody SED with temperature $T_{BB} = 4 \times 10^{5}$ K. At this temperature there are comparable amounts of UV and X-ray photons with $f_{UV} \sim 0.15$ and $f_X \sim 0.85$. We then study the effects of changing SED by introducing an absorption opacity $0 \leq \kappa_a / \kappa_{es} \leq 3$ in either the UV or X-ray bands. Absorption along the wind leads to a radial dependence of the blackbody temperature and hence changes in the force multiplier. It also leads to deviations from the $1/r^2$ scaling of the radiation flux. 

In Fig \ref{fig:dynamical_summary} we show the radial dependence of density, velocity, blackbody temperature and mass flux after the flow has reached a steady state solution. The dashed lines are for models with $\kappa_a = 0$ with irradiating blackbody spectra with temperatures $T_{BB} [K] = 3 \times 10^{5}$ (red), $3.5 \times 10^{5}$ (orange), $4 \times 10^{5}$ (green), $4.5 \times 10^{5}$ (cyan) and $5 \times 10^{5}$ (blue). The solid lines show models with $\kappa_{a,UV} = 3 \kappa_{es}$ (blue, solid line) and $\kappa_{a,X} = 3 \kappa_{es}$ (red, solid line). The colors for these models are chosen to match the optically thin spectra with approximately the same blackbody temperature at the outer boundary. 

First, we consider the trends in the models with no absorption (dashed lines). Decreasing the temperature of the irradiating blackbody spectra leads to an increase in the mass flux. As the temperature is decreased, the spectra becomes softer and the UV fraction $f_{UV}$ increases. The softer spectra have stronger force multiplier and more UV flux. The UV band tends to have a higher force multiplier $M_{UV} \sim 10$, compared to $M_{X} \sim 1$ in the outer parts of the wind. The enhanced mass flux is due to both an enhanced density and velocity.

Now consider the models with absorption, with SED temperature $T_{BB} = 4 \times 10^5$ K at the inner boundary. The wind column has a total optical depth $\tau \approx 0.5$, so the flow is marginally optically thick. The absorption is sufficient to change the effective blackbody temperature of the spectra by $\Delta T_{BB} \approx 1 \times 10^{5}$ K. At the outer boundary, this corresponds to a UV fraction $f_{UV} \approx 0.05$ or a change of approximately 30\%. We plot the solution in the same color as the optically thin model with the same SED at the outer boundary.

In the case of UV absorption, the wind profile closely matches the optically thin wind with $T_{BB} = 3 \times 10^5$ K (blue dashed line). In contrast, the X-ray absorption model does not match the $T_{BB} = 5 \times 10^5$ K very well (red dashed line). In fact, it is in good agreement with the $T_{BB} = 3.5 \times 10^5$ K model, for the mass flux and density profiles, though there is some discrepancy between the velocity profiles.

One can understand these results by considering the radiation force on the flow. In Fig \ref{fig:Mt_avg} we plot $M_{UV}$ as a function of radial position for a sample of runs. The radiation force is a product of the irradiating force and the force multiplier. Since lines in the UV are much stronger than those in the X-rays, with $M_{UV}/M_{X} \sim 10$, we can approximate the total radiation force as 
\begin{equation}
    f_{rad} \sim F_{UV} M_{UV}(f_{UV}).
    \label{eq:frad}
\end{equation}
In the case of the wind with X-ray absorption, the UV flux profile is the same as in the fiducial case - UV photons are unattenuated. In addition, we see from Fig \ref{fig:Mt_avg}-\ref{fig:Mt_avg_summary} that the force multiplier profiles nearly follow each other as well. Thus the wind dynamics are nearly identical. In contrast, the case with UV attenuation is less straight forward. Firstly, there is UV attenuation so there is a drop in the UV flux i.e. the first term in equation (\ref{eq:frad}). This is however compensated for by a nearly two times stronger force multiplier, due to the softer SED. Thus the attenuated UV solution is a hybrid one, with mass flux between the $T = 3 \times 10^5 \ \rm{K}$ and $4 \times 10^5 \ \rm{K}$ models but a faster velocity profile than both of them. In fact, the mass flux is lower than the unattenuated model because the mass flux is determined by the dynamics interior to the critical point and we see the force multiplier weaker for $r - r_* \lesssim 1.05$. Further out in the wind, the force multiplier is stronger, hence allowing for the velocity to become larger. 

\subsection{Full Velocity Gradient}

\begin{figure*}
    \centering
    \includegraphics[scale=0.8]{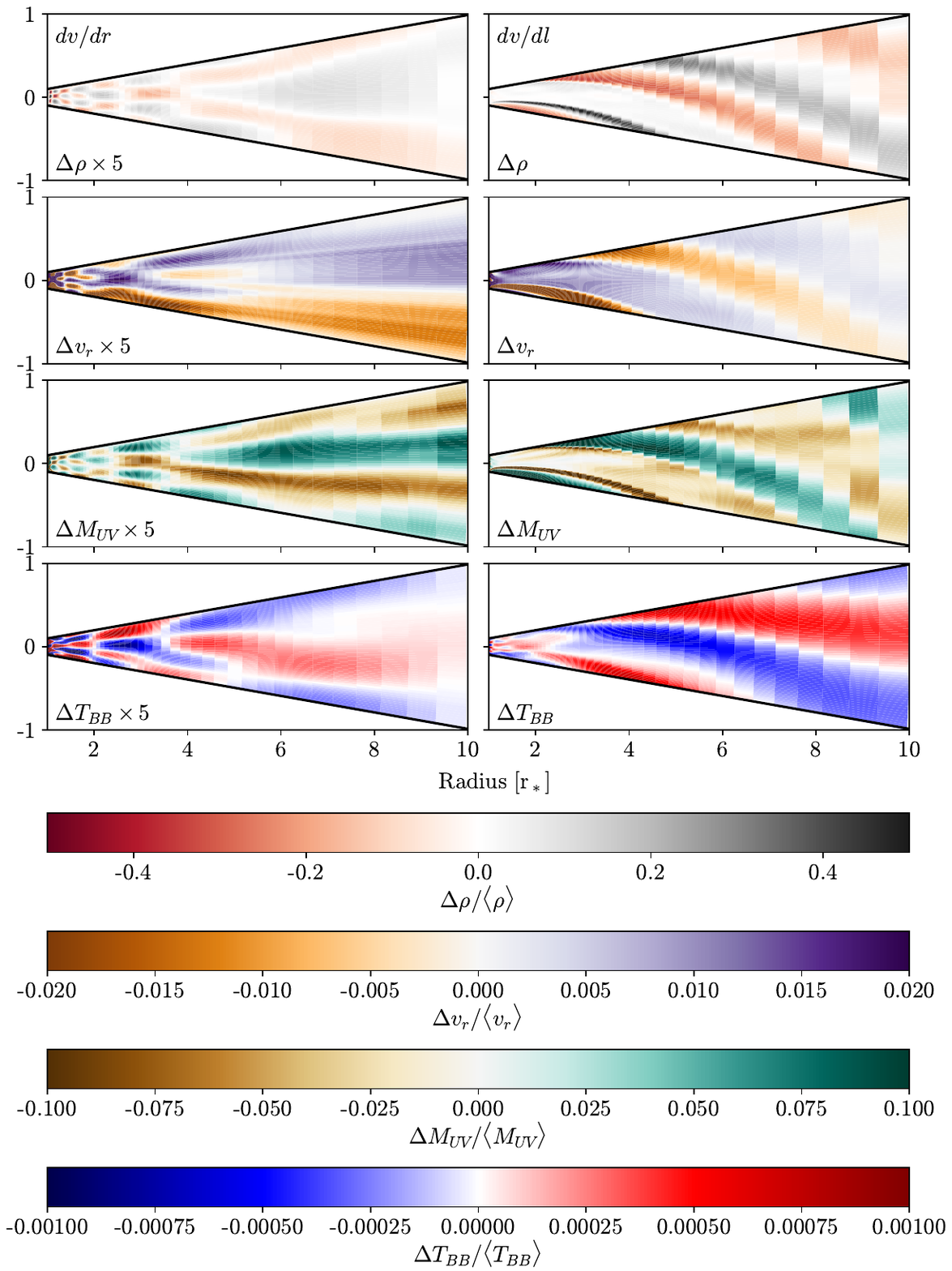}
    \caption{ Deviations in the azimuthal average for density, radial velocity, UV contribution to the force multiplier and SED temperature for models with $\kappa_{a,UV} = 3 \kappa_{es}$ using the radial streaming approximation (left panels) and full velocity gradient (right panel). To use the same color bars we have magnified the size of the deviations by 5x in the radial streaming approximation in these plots. }
    \label{fig:perturbations_summary}
\end{figure*}

We next consider models where we use the full velocity gradient $d\mathbf{v}/d\mathbf{l}$ in equation (\ref{eq:dvdl}). We consider models with the same parameters as described in the previous section.

All solutions reach an azimuthally averaged steady state. Compared to models that use the radial streaming approximation, winds tend to carry a higher mass flux due to an enhanced outflow velocity. This occurs because the line force can be effectively larger since rays with a small azimuthal component can have a lower optical depth parameter than purely radial rays and provide a larger impulse.

We study deviations from the azimuthal average in these wind solutions. In particular, we consider
\begin{equation}
    \Delta X = X - \langle X\rangle
\end{equation}
where $\langle X\rangle$ is the azimuthal average of the quantity X. For solutions with the fastest winds, $\kappa_X = 3$ and $t \sim dv/dl$ and $t \sim dv/dr$, deviations form at early times but these are advected away at late times and do not persist. For solutions where the wind is less fast, $\kappa_{UV}, \kappa_{X} \leq 1$, small inhomoneities in the density and velocity persist to late times but $\delta \rho, \delta v_r, \delta T_{BB} \lesssim 0.1\%$ and reach a quasi-steady state.

Thus we find for azimuthal perturbations to be large ($\Delta X / \langle X\rangle \gtrsim 0.1$) and persistent we must meet two conditions 1) the line force must be strong. The enhanced strength can be due to using the full velocity gradient or UV absorption is large. 2) the wind must not be too fast, or perturbatons are advected away. We find several instances of winds which satisfy these criteria, and each exhibits different correlations in their perturbations.

When $\kappa_{UV} = 0$, the inhomogenities are well correlated, with $\Delta \rho \sim \Delta v_r \sim \Delta M_{UV}$. This indicates that the mechanism for growing the perturbations does not rely on variations in the local SED. By contrast cases with $\kappa_{UV} = 3$, with either form of the velocity gradient, exhibit azimuthal variations in their local blackbody temperature. In Fig \ref{fig:perturbations_summary} we plot the azimuthal perturbations in $\Delta \rho$, $\Delta v_r$, $\Delta M_{UV}$ and $\Delta T_{BB}$ for these two models. Both models generate azimuthal perturbations. In the $dv/dr$ case, perturbations are weaker so we have magnified them by a factor of 5x to use the same colorbars. With the radial streaming approximation, there is no or a weak correlation between $\Delta v_r$ and $\Delta \rho$ and $\Delta M_{UV}$. In the $dv/dl$ case, all perturbations appear correlated to one another. In fact, the wind appears to carry a spiral density wave propagating radially out of the flow. These results suggest that the formation of azimuthal structures is more a function of how the velocity gradient is computed. Including the effects of a local SED on the force multiplier calculation does not increase the formation of these structures in the models we have investigated.

\section{Discussion}
\label{sec:discussion}

We have developed a method to account for a changing SED within a hydrodynamics simulation and apply this change to the treatment of the gas microphysics, in real time. We have applied this study to line driven winds and explicitly demonstrated that changes to the SED alter the radiation/gas coupling via the force multiplier and alter the hydrodynamics. In particular, outflows are suppressed when UV photons, which are the dominant contributor to line driving, are absorbed via $\kappa_a$. The suppression of the wind cannot be understood only through a decrease in the available momentum in the radiation field, but also due to a change in the gas/radiation coupling via $M_{UV}$.   

We further studied the formation of azimuthal instabilities in the wind. Cases where the radiation force was relatively weak failed to produce noticeable asymmetries. Winds that were sufficiently fast produced inhomogeneities at early time, but these were advected away from the star at late times. When the line force was sufficiently strong, inhomogeneities form and continue evolving at late times. Our results suggest that though deviations in the local irradiating SED can develop, these do not directly lead to the formation and growth of density and velocity perturbations. The fractional change in the blackbody temperature is $\lesssim 10^{-3}$ and this change is too small to induce perturbations in the hydrodynamical variables.  Rather we find that growth due to a stronger line force because of shadowing by density features, larger absorption opacity or the use of the full velocity gradient are sufficient for inhomogeneities to grow. 

An immediate application of these methods is to the study of line driven disc winds, such as CVs and AGNs. Typically, accretion discs are treated as multi-temperature blackbodies. This formalism allows for a position dependent force multiplier to be computed from the local SED. In the context of AGN, assuming an optically thin wind, the local SED is highly positionally dependent \cite{Smith24}. Further, line driven disc winds are highly unstable, both in terms of their gross wind properties \citep{DDP24} and their fine scale structure \citep{DP2018a,DP2018b}. These results suggest the need for a force multiplier prescription that depends on the \emph{local} irradiating SED. Such a treatment may be particularly important for studying density inhomogenieties, so called clumps, which may shadow further outlying gas in the flow.

A further application is in the context of massive star winds, particularly Wolf-Rayet (WR) stars. WR winds are thought to be radiatively driven, but crucially have optically thick winds. Further, observations indicate the presence of X-rays in the spectra, though these are not expected to originate from the corona due to the temperature being too low. It remains an open problem to determine the source of the X-rays, with potential solutions including shocks within the wind, magnetically confined shocks, non-thermal X-ray emission, or interaction with a companion or the ISM \citep{2016AdSpR..58..739O}. As we have shown, the X-ray emission may affect the force multiplier, and hence the radiation force, altering the wind.  

We have compiled force multiplier tables for blackbody SEDs, but one could generalize this method to other spectra. The blackbody choice is a convenient one, as it can be described by a single parameter. A limitation of this method, that might be improved upon, is that it uses only two frequency bands. For additional bands modeled in the hydrodynamics, one also has to model an equally larger parameter space with the photionization code. For instance, one could choose an SED consisting of a blackbody and a powerlaw tail. One could then extract the blackbody temperature from $f_{UV}$, and any additional X-ray photons interpreted as coming from the blackbody tail. Such a two parameter SED may be applied to studying AGN coronae for example. Given that we already want to model the optical depth parameter of the gas (or in the case of gas heating  the gas temperature), it may not be practical to go far beyond two SED parameters.

\appendix

\section{Flux Averaged Force Multiplier}
\label{sec:FluxAvgMt}

\begin{figure*}
    \centering
    \includegraphics[scale=0.8]{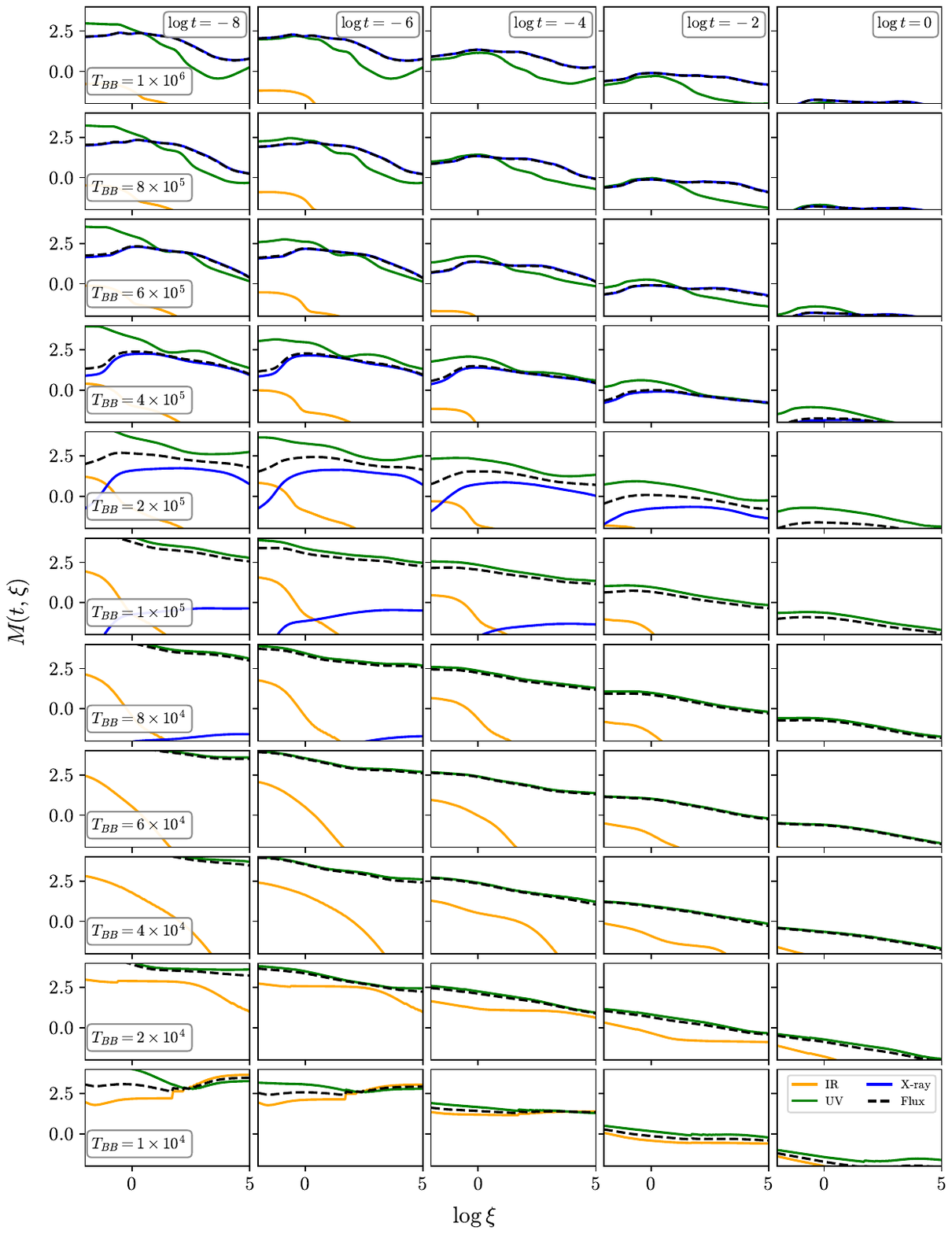}    
    \caption{Force multiplier for IR (orange), UV (green line) and X-ray (blue line) bands as a function of ionization parameter for various optical depth parameters for gas irradiated by a blackbody SED with temperatures in the range $10^4 \ \rm{K} \leq T_{BB} \leq 10^6 \ \rm{K}$. The dashed black line shows the flux weighted force multiplier for a blackbody spectrum at the same temperature.}
    \label{fig:Mt_avg}
\end{figure*}

\begin{figure*}
    \centering
    \includegraphics[scale=0.8]{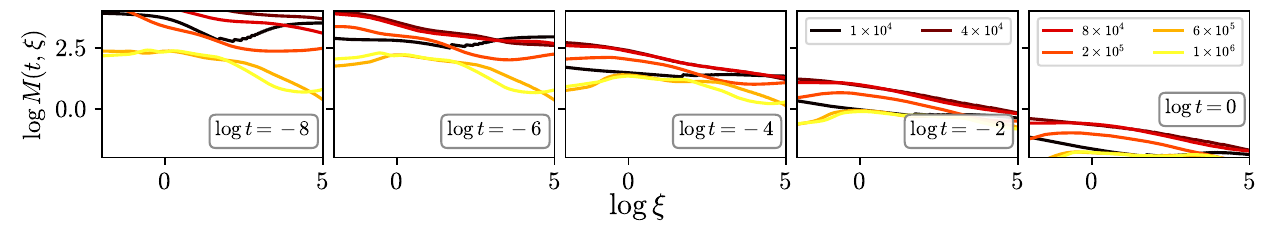} 
    \caption{Summary of the flux weighted force multiplier for blackbody SEDs with temperature range $10^4 \ \rm{K} \leq T_{BB} \leq 10^6$ K.}
    \label{fig:Mt_avg_summary}
\end{figure*}

In Fig \ref{fig:Mt_avg} we plot the contributions to the force multiplier for the IR (orange line), UV (green line) and X-ray (blue line) bands, as well as the flux weighted average for a blackbody spectrum at the same temperature (black line) as a function of ionization parameter for various optical depth parameters for gas irradiated by blackbodies in the temperature range $10^{4} \ \rm{K} \leq T_{BB} \leq 10^{6} \ \rm{K}$. Fig \ref{fig:Mt_avg_summary} shows a summary of the flux weighted force multipliers for each of these temperatures.

At low temperatures, $T \lesssim 2 \times 10^{5}$ K, the spectrum is dominated by UV photons, hence the flux averaged force multiplier nearly matches the UV force multiplier. Likewise, at high temperatures, $T \gtrsim 8 \times 10^5$ K, the force multiplier is dominated by X-ray band contributions. At intermediate temperatures, both bands contribute to the force multiplier, with UV contributions tending to dominate at the lowest ionizations and X-rays contributing at higher ionizations. 

As a test of our numerical methods we investigated stellar line driven wind models with either 1 (grey band) or 2 (UV + X-ray) radiation bands in the case of an optically thin wind. In this case, the irradiating SED is constant throughout the wind so the grey approximation is consistent provided we use the flux averaged force multipliers 
\begin{equation}
    M_{\rm{flux}} = f_{IR} M_{IR} + f_{UV} M_{UV} + f_{X} M_{X}
\end{equation}
where $f_{i}$ are the band fractions at the irradiating SED blackbody temperatures. We found that both methods yield the same wind solutions to numerical accuracy, validating our radiation transfer method and force multiplier interpolation method.

\section{Spectral Averaged Force Multiplier}
\label{sec:SpecAvg}

\begin{figure*}
    \centering
    \includegraphics[scale=0.52]{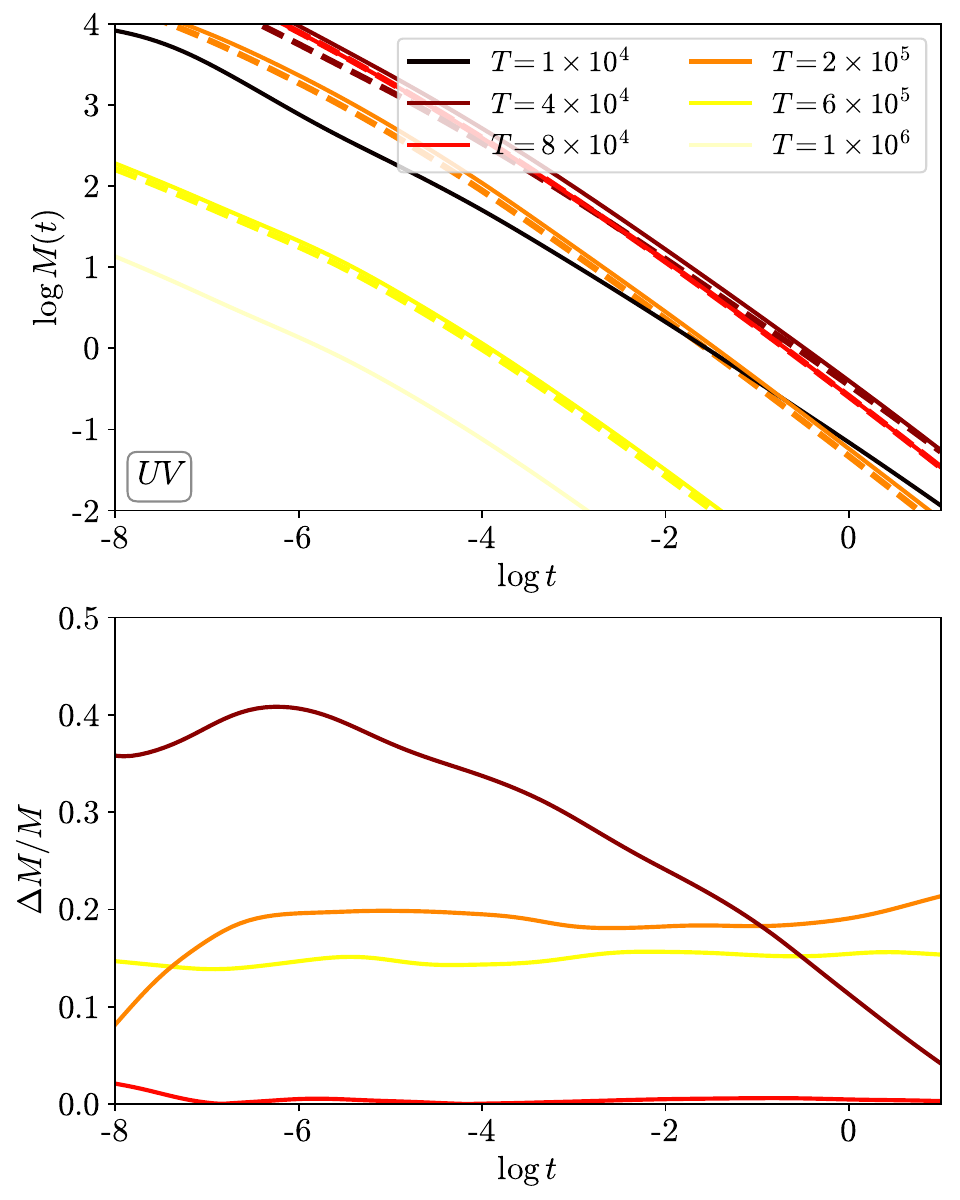}
    \includegraphics[scale=0.52]{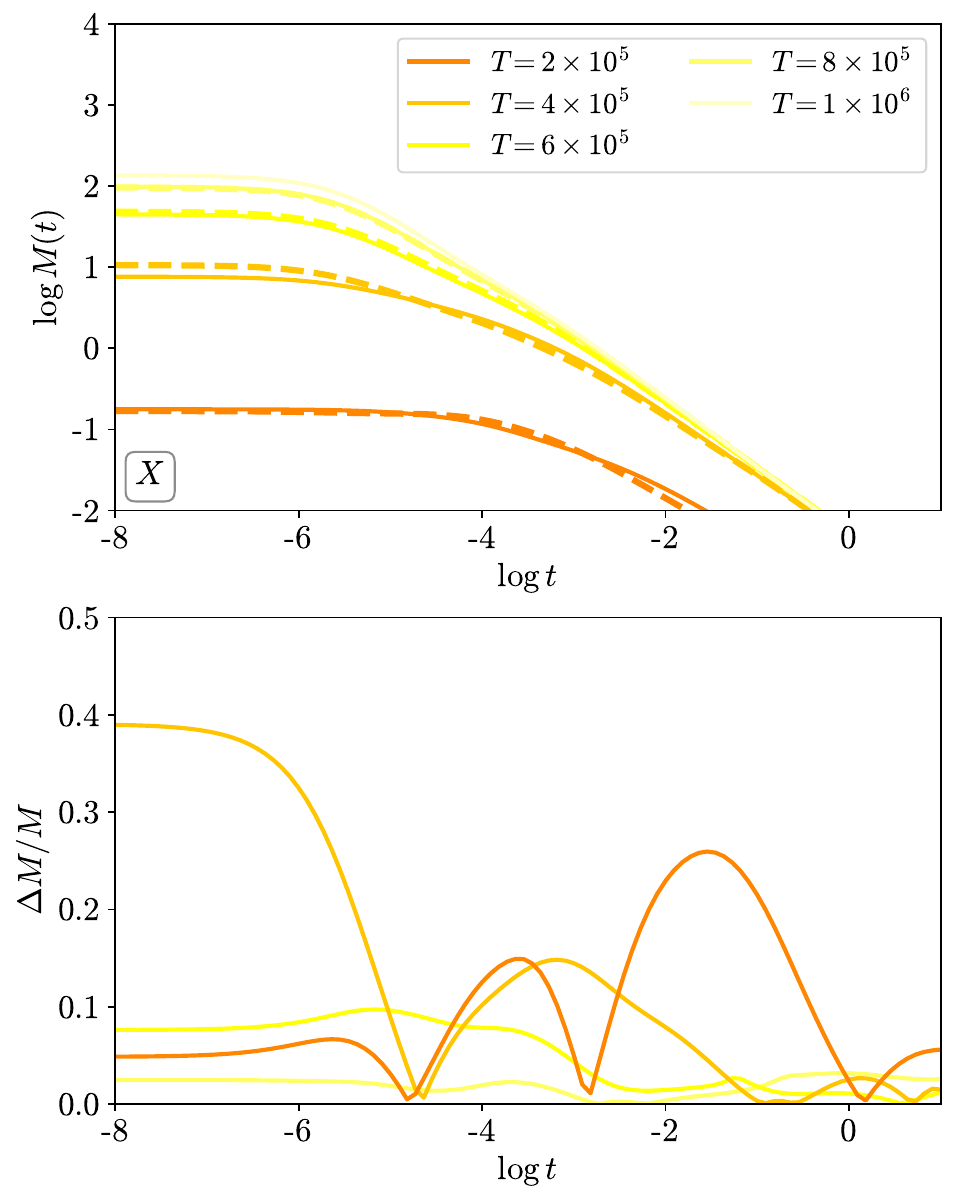}
    \caption{Force multiplier for UV (left panels) and X-ray (right panels) bands. \textit{Top panels} Force multiplier contributions from the relevant band for an irradiating blackbody of the indicated temperature. The solid lines show the results of the \textsc{XSTAR} modeling, whereas the dashed line is the fit with that same temperature removed from the data set. \textit{Bottom panels} Relative error $\Delta M/ M$ of the force multiplier for each irradiating SED.}
    \label{fig:Mt_int}
\end{figure*}

The method described in Appendix \ref{sec:FluxAvgMt} allows one to account for multiple frequency bands propagating through the gas. However, the force multiplier prescription implicitly assumed the gas is irradiated by a fixed SED. As the hydrodynamics simulation evolves, the corresponding SED irradiating a particular cell may change and hence the assumptions used to compute the force multiplier may no longer be consistent.

To alleviate this limitation, we compile all force multiplier data as a function of UV fraction $f_{UV}$. One can think of this as instead of indexing our data by the blackbody temperature, we instead index it in terms of $f_{UV}$. If we were to restrict ourselves to two frequency bands, this is certainly possible i.e blackbodies are described by a single parameter. However, we can in fact add an additional third band, IR, since in our range of blackbody temperatures only two bands contribute $\gtrsim 1\%$ to the spectrum. Our interpolation scheme thus effectively has two regions, a ``cold'' SED where $f_{IR} = 1-f_{UV}$ and a ``hot'' SED where $f_{X} = 1-f_{UV}$. We determine which scheme to use by comparing the energy densities $J_{IR}$ and $J_{X}$. In each of these ranges, $f_{UV}$ is a monotonic function of blackbody temperature, so we can interpolate the force multiplier data and produce a function $M(t,f_{UV})$.

We test our interpolation scheme by removing a blackbody temperature from the full data set, interpolating over the remaining temperatures and comparing the resulting interpolated force multiplier with the data from the removed blackbody temperature. In Fig \ref{fig:Mt_int} we show the result of this test for the UV (left panels) and X-ray (right panels) contributions to the force multiplier. The top panels show the force multiplier contributions to the relevant band for an irradiating blackbody of the indicated temperature. The solid lines show the results of the \textsc{XSTAR} modeling, whereas the dashed line is the interpolation scheme derived by excluding that temperature from the data set. We do this for every temperature except for the extrema, where the fit is bad since there is no derivative information. The bottom panels show the relative error $\Delta M/ M$ of the force multiplier for each irradiating SED. 

First, we note that the fit for $\log M$ is pretty good, most likely owing to the fact that the curves are monotonic with blackbody temperature. However, since we are fitting a powerlaw, the absolute fit still shows errors of $\sim 10s \%$. It may be possible to improve on this fitting by providing a more uniform coverage of photoionization models in $f_{UV}$ space. Though the models cover the blackbody temperature space relatively uniformally, the Planck function is highly non-linear, resulting in large jumps in the radiation fraction in each band.

\section*{Acknowledgments} 
Support for this work was provided by the National Aeronautics and Space Administration under TCAN grant 80NSSC21K0496. We thank Randal Dannen for having made the XSTAR models used in this work available. We thank Tim Kallman, Daniel Proga, Yan-Fei Jiang and the entire DAWN TCAN collaboration for fruitful discussions.  The authors acknowledge Research Computing at The University of Virginia for providing computational resources and technical support that have contributed to the results reported within this publication.

\section*{Data Availability Statement}
The simulations were performed with the publicly available code \textsc{Athena++} available at https://github.com/PrincetonUniversity/athena and \textsc{XSTAR} available at https://heasarc.gsfc.nasa.gov/xstar/xstar.html The authors will provide the force multiplier tables to interested parties. The authors will also provide any additional problem generators and input files upon request.


\bibliographystyle{mnras}
\bibliography{progalab-shared}

\begin{thebibliography}{}
\makeatletter
\relax
\def\mn@urlcharsother{\let\do\@makeother \do\$\do\&\do\#\do\^\do\_\do\%\do\~}
\def\mn@doi{\begingroup\mn@urlcharsother \@ifnextchar [ {\mn@doi@}
  {\mn@doi@[]}}
\def\mn@doi@[#1]#2{\def\@tempa{#1}\ifx\@tempa\@empty \href
  {http://dx.doi.org/#2} {doi:#2}\else \href {http://dx.doi.org/#2} {#1}\fi
  \endgroup}
\def\mn@eprint#1#2{\mn@eprint@#1:#2::\@nil}
\def\mn@eprint@arXiv#1{\href {http://arxiv.org/abs/#1} {{\tt arXiv:#1}}}
\def\mn@eprint@dblp#1{\href {http://dblp.uni-trier.de/rec/bibtex/#1.xml}
  {dblp:#1}}
\def\mn@eprint@#1:#2:#3:#4\@nil{\def\@tempa {#1}\def\@tempb {#2}\def\@tempc
  {#3}\ifx \@tempc \@empty \let \@tempc \@tempb \let \@tempb \@tempa \fi \ifx
  \@tempb \@empty \def\@tempb {arXiv}\fi \@ifundefined
  {mn@eprint@\@tempb}{\@tempb:\@tempc}{\expandafter \expandafter \csname
  mn@eprint@\@tempb\endcsname \expandafter{\@tempc}}}

\bibitem[\protect\citeauthoryear{{Blondin}}{{Blondin}}{1994}]{Blondin1994}
{Blondin} J.~M.,  1994, \mn@doi [\apj] {10.1086/174853}, \href
  {https://ui.adsabs.harvard.edu/abs/1994ApJ...435..756B} {435, 756}

\bibitem[\protect\citeauthoryear{{Blondin}, {Kallman}, {Fryxell}  \&
  {Taam}}{{Blondin} et~al.}{1990}]{Blondin1990}
{Blondin} J.~M.,  {Kallman} T.~R.,  {Fryxell} B.~A.,   {Taam} R.~E.,  1990,
  \mn@doi [\apj] {10.1086/168865}, \href
  {https://ui.adsabs.harvard.edu/abs/1990ApJ...356..591B} {356, 591}

\bibitem[\protect\citeauthoryear{{Dannen} \& {Proga}}{{Dannen} \&
  {Proga}}{2023}]{Dannen2023}
{Dannen} R.,  {Proga} D.,  2023, in American Astronomical Society Meeting
  Abstracts. p. 360.46

\bibitem[\protect\citeauthoryear{{Dannen}, {Proga}, {Kallman}  \&
  {Waters}}{{Dannen} et~al.}{2019}]{Dannen19}
{Dannen} R.~C.,  {Proga} D.,  {Kallman} T.~R.,   {Waters} T.,  2019, \mn@doi
  [\apj] {10.3847/1538-4357/ab340b}, \href
  {https://ui.adsabs.harvard.edu/abs/2019ApJ...882...99D} {882, 99}

\bibitem[\protect\citeauthoryear{{Dannen}, {Proga}, {Waters}  \&
  {Dyda}}{{Dannen} et~al.}{2020}]{Dannen20}
{Dannen} R.~C.,  {Proga} D.,  {Waters} T.,   {Dyda} S.,  2020, \mn@doi [\apjl]
  {10.3847/2041-8213/ab87a5}, \href
  {https://ui.adsabs.harvard.edu/abs/2020ApJ...893L..34D} {893, L34}

\bibitem[\protect\citeauthoryear{{Dannen}, {Proga}, {Waters}  \&
  {Dyda}}{{Dannen} et~al.}{2024}]{Dannen2024}
{Dannen} R.,  {Proga} D.,  {Waters} T.,   {Dyda} S.,  2024, \mn@doi [\apj]
  {10.3847/1538-4357/ad0da5}, \href
  {https://ui.adsabs.harvard.edu/abs/2024ApJ...961..221D} {961, 221}

\bibitem[\protect\citeauthoryear{{Dyda} \& {Proga}}{{Dyda} \&
  {Proga}}{2018a}]{DP2018a}
{Dyda} S.,  {Proga} D.,  2018a, \mn@doi [\mnras] {10.1093/mnras/sty030}, \href
  {https://ui.adsabs.harvard.edu/abs/2018MNRAS.475.3786D} {475, 3786}

\bibitem[\protect\citeauthoryear{{Dyda} \& {Proga}}{{Dyda} \&
  {Proga}}{2018b}]{DP2018b}
{Dyda} S.,  {Proga} D.,  2018b, \mn@doi [\mnras] {10.1093/mnras/sty1257}, \href
  {https://ui.adsabs.harvard.edu/abs/2018MNRAS.478.5006D} {478, 5006}

\bibitem[\protect\citeauthoryear{{Dyda}, {Dannen}, {Waters}  \& {Proga}}{{Dyda}
  et~al.}{2017}]{Dyda17}
{Dyda} S.,  {Dannen} R.,  {Waters} T.,   {Proga} D.,  2017, \mn@doi [\mnras]
  {10.1093/mnras/stx406}, \href
  {https://ui.adsabs.harvard.edu/abs/2017MNRAS.467.4161D} {467, 4161}

\bibitem[\protect\citeauthoryear{{Dyda}, {Davis}  \& {Proga}}{{Dyda}
  et~al.}{2024}]{DDP24}
{Dyda} S.,  {Davis} S.~W.,   {Proga} D.,  2024, \mn@doi [\mnras]
  {10.1093/mnras/stae1159}, \href
  {https://ui.adsabs.harvard.edu/abs/2024MNRAS.530.5143D} {530, 5143}

\bibitem[\protect\citeauthoryear{{Gonz{\'a}lez}, {Audit}  \&
  {Huynh}}{{Gonz{\'a}lez} et~al.}{2007}]{2007A&A...464..429G}
{Gonz{\'a}lez} M.,  {Audit} E.,   {Huynh} P.,  2007, \mn@doi [\aap]
  {10.1051/0004-6361:20065486}, \href
  {https://ui.adsabs.harvard.edu/abs/2007A&A...464..429G} {464, 429}

\bibitem[\protect\citeauthoryear{{Higginbottom}, {Proga}, {Knigge}, {Long},
  {Matthews}  \& {Sim}}{{Higginbottom} et~al.}{2014}]{Higginbottom14}
{Higginbottom} N.,  {Proga} D.,  {Knigge} C.,  {Long} K.~S.,  {Matthews} J.~H.,
    {Sim} S.~A.,  2014, \mn@doi [\apj] {10.1088/0004-637X/789/1/19}, \href
  {https://ui.adsabs.harvard.edu/abs/2014ApJ...789...19H} {789, 19}

\bibitem[\protect\citeauthoryear{{Higginbottom}, {Knigge}, {Sim}, {Long},
  {Matthews}, {Hewitt}, {Parkinson}  \& {Mangham}}{{Higginbottom}
  et~al.}{2020}]{Higginbottom2020}
{Higginbottom} N.,  {Knigge} C.,  {Sim} S.~A.,  {Long} K.~S.,  {Matthews}
  J.~H.,  {Hewitt} H.~A.,  {Parkinson} E.~J.,   {Mangham} S.~W.,  2020, \mn@doi
  [\mnras] {10.1093/mnras/staa209}, \href
  {https://ui.adsabs.harvard.edu/abs/2020MNRAS.492.5271H} {492, 5271}

\bibitem[\protect\citeauthoryear{{Jiang}}{{Jiang}}{2022}]{Jiang2022}
{Jiang} Y.-F.,  2022, \mn@doi [\apjs] {10.3847/1538-4365/ac9231}, \href
  {https://ui.adsabs.harvard.edu/abs/2022ApJS..263....4J} {263, 4}

\bibitem[\protect\citeauthoryear{{Jiang}, {Stone}  \& {Davis}}{{Jiang}
  et~al.}{2012}]{Jiang12}
{Jiang} Y.-F.,  {Stone} J.~M.,   {Davis} S.~W.,  2012, \mn@doi [\apjs]
  {10.1088/0067-0049/199/1/14}, \href
  {https://ui.adsabs.harvard.edu/abs/2012ApJS..199...14J} {199, 14}

\bibitem[\protect\citeauthoryear{{Kallman} \& {Bautista}}{{Kallman} \&
  {Bautista}}{2001}]{XSTAR2001}
{Kallman} T.,  {Bautista} M.,  2001, \mn@doi [\apjs] {10.1086/319184}, \href
  {https://ui.adsabs.harvard.edu/abs/2001ApJS..133..221K} {133, 221}

\bibitem[\protect\citeauthoryear{{Lamers} \& {Cassinelli}}{{Lamers} \&
  {Cassinelli}}{1999}]{1999isw..book.....L}
{Lamers} H. J.~G.~L.~M.,  {Cassinelli} J.~P.,  1999, {Introduction to Stellar
  Winds}

\bibitem[\protect\citeauthoryear{{Levermore} \& {Pomraning}}{{Levermore} \&
  {Pomraning}}{1981}]{1981ApJ...248..321L}
{Levermore} C.~D.,  {Pomraning} G.~C.,  1981, \mn@doi [\apj] {10.1086/159157},
  \href {https://ui.adsabs.harvard.edu/abs/1981ApJ...248..321L} {248, 321}

\bibitem[\protect\citeauthoryear{{Matthews} et~al.,}{{Matthews}
  et~al.}{2025}]{Matthews2025}
{Matthews} J.~H.,  et~al., 2025, \mn@doi [\mnras] {10.1093/mnras/stae2677},
  \href {https://ui.adsabs.harvard.edu/abs/2025MNRAS.536..879M} {536, 879}

\bibitem[\protect\citeauthoryear{{Oskinova}}{{Oskinova}}{2016}]{2016AdSpR..58..739O}
{Oskinova} L.~M.,  2016, \mn@doi [Advances in Space Research]
  {10.1016/j.asr.2016.06.030}, \href
  {https://ui.adsabs.harvard.edu/abs/2016AdSpR..58..739O} {58, 739}

\bibitem[\protect\citeauthoryear{{Proga}, {Stone}  \& {Kallman}}{{Proga}
  et~al.}{2000}]{PSK2000}
{Proga} D.,  {Stone} J.~M.,   {Kallman} T.~R.,  2000, \mn@doi [\apj]
  {10.1086/317154}, \href
  {https://ui.adsabs.harvard.edu/abs/2000ApJ...543..686P} {543, 686}

\bibitem[\protect\citeauthoryear{{Robinson}, {Avestruz}  \&
  {Gnedin}}{{Robinson} et~al.}{2024}]{2024arXiv240619446R}
{Robinson} D.,  {Avestruz} C.,   {Gnedin} N.~Y.,  2024, \mn@doi [arXiv
  e-prints] {10.48550/arXiv.2406.19446}, \href
  {https://ui.adsabs.harvard.edu/abs/2024arXiv240619446R} {p. arXiv:2406.19446}

\bibitem[\protect\citeauthoryear{{Smith}, {Proga}, {Dannen}, {Dyda}  \&
  {Waters}}{{Smith} et~al.}{2024}]{Smith24}
{Smith} K.,  {Proga} D.,  {Dannen} R.,  {Dyda} S.,   {Waters} T.,  2024,
  \mn@doi [\apj] {10.3847/1538-4357/ad4a70}, \href
  {https://ui.adsabs.harvard.edu/abs/2024ApJ...970..150S} {970, 150}

\bibitem[\protect\citeauthoryear{{Stone} \& {Proga}}{{Stone} \&
  {Proga}}{2009}]{Stone2009}
{Stone} J.~M.,  {Proga} D.,  2009, \mn@doi [\apj]
  {10.1088/0004-637X/694/1/205}, \href
  {https://ui.adsabs.harvard.edu/abs/2009ApJ...694..205S} {694, 205}

\makeatother
\end{thebibliography}

\bsp	
\label{lastpage}

\end{document}